\documentclass[pra,twocolumn,showpacs,amsmath,amssymb,floatfix,superscriptaddress]{revtex4}

\usepackage{graphicx}
\usepackage{color}
\usepackage{amsmath}
\usepackage{amsfonts}

\begin{document}

\title{Two coupled nonlinear cavities in a driven-dissipative environment}

\author{Bin Cao}
\thanks{These authors contributed equally}
\affiliation{Joint Quantum Institute, NIST/University of Maryland, College Park, MD 20742, USA}

\author{Khan W. Mahmud}
\thanks{These authors contributed equally}
\affiliation{Joint Quantum Institute, NIST/University of Maryland, College Park, MD 20742, USA}

\author{Mohammad Hafezi}
\affiliation{Joint Quantum Institute, NIST/University of Maryland, College Park, MD 20742, USA}
\affiliation{Kavli Institute of Theoretical Physics, Santa Barbara, CA 93106, USA}
\affiliation{Department of Electrical and Computer Engineering and Institute for Research in Electronics and Applied Physics,
University of Maryland, College Park, MD 20742, USA}

\begin{abstract}
We investigate two coupled nonlinear cavities that are coherently driven in a dissipative environment. We perform semiclassical, numerical and analytical quantum studies of this dimer model when both cavities are symmetrically driven. In the semiclassical analysis, we find steady-state solutions with different photon occupations in two cavities. Such states can be considered analogs of the closed system double well symmetry breaking states. We analyze the occurrence and properties of these localized states in the system parameter space and examine how the symmetry breaking states, in form of a bistable pair, are associated to the single cavity bistable behavior. In a full quantum calculation of the master equation dynamics that includes quantum fluctuations, the symmetry breaking states and bistability disappear due to the quantum fluctuations. In quantum trajectory picture, we observe enhanced quantum jumps and switching which indicate the presence of the underlying semiclassical symmetry breaking states. Finally, we present a set of analytical solutions for the steady state correlation functions using the complex P-representation and discuss its regime of validity.
\end{abstract}


\maketitle

\section{Introduction}

Experimental advances in generating strongly interacting photons opened the opportunity to explore various many-body quantum phenomena in a new context~\cite{carusotto13,hartmann16,angelakis16}. Photonic systems are unique in the sense that it is naturally an open quantum system where the adding and destroying of photons can be accomplished by drive and dissipation in a controlled way. Thus it is an ideal system to study many outstanding questions on open systems dynamics, dissipative phase transitions~\cite{houck16,carmichael15} and the effects of interactions in a dissipative environment. The simplest model to study strongly interacting bosons on a lattice is the celebrated Bose-Hubbard model where atoms have on-site interactions and can hop across lattices~\cite{fisher89,greiner01}. There have been several recent proposals~\cite{hartmann06,greentree06,angelakis07,hartmann10,hafezi15} on achieving this model with photons and open systems, such as with photons in coupled cavity arrays, superconducting circuit QED and polaritons. With added drive and dissipation, the Bose-Hubbard model does not exhibit superfluid or Mott insulator phases but gives rise to mixed state nonequilibrium steady states and phases~\cite{ryan16,ciuti13,ciuti14,jaksch15}, the nature and generation of which are not fully understood.

A starting point for understanding the complex dynamics of driven dissipative photonic cavity arrays can be a two-site model. Such a two-site driven dissipative nonlinear model may be realized with systems such as two coupled photon cavities~\cite{bloch16}, circuit QED systems~\cite{houck14} and two coupled micropillars~\cite{jbloch12}. This system is also referred to as a photonic molecule~\cite{molecule04}, dimer or a double well. For the closed system, the physics of double well has been studied in great detail~\cite{oberthaler05,smerzi97,mahmud02,mahmud05,jbloch13,trenkwalder15}, giving rise to phenomena such as the Josephson effect, matter wave interference and self-trapped and symmetry breaking states, among others. For open systems, studies of two coupled cavities have appeared in several contexts in both theory~\cite{olsen06,sarchi08,tureci10,houck10,liew10,ciuti11,lemande14,grujic13,campbell16} and experiment~\cite{bloch16,houck14,hamel14,bloch16b}. For an end-driven cavity, studies focused on the topics of unconventional photon blockade~\cite{liew10,ciuti11,lemande14} and multi-stability~\cite{bloch16}, among others. For symmetrically driven cavities, studies appeared on quantum correlations~\cite{tureci10}, classical to quantum phase transitions for a Jaynes-Cummings dimer~\cite{houck14}, and symmetry breaking for incoherent drives~\cite{hamel14}, among others.

In this article, we explore the physics of two coupled nonlinear cavities in a dissipative setting where both sites are driven coherently and symmetrically. We perform semiclassical and quantum analysis of the system investigating the complex interplay of many competing terms such as hopping, interaction, drive, dissipation and detuning. In a semiclassical treatment, we show that the nonequilibrium steady states have asymmetric number density in the two cavities in addition to the expected symmetry preserving states. These states are the driven-dissipative analog of the closed system double well symmetry breaking or self-trapped states~\cite{oberthaler05,smerzi97,mahmud02} with a fundamental difference that these are not minimum energy states but long-time steady states resulting from the competition in drive and dissipation. These can be understood from the bistability of a single driven cavity; when two cavities are coupled with small but nonzero tunneling, the low density and high density bistable branches of a single cavity are hybridized to form two symmetry breaking steady states with unequal photon occupations in the two cavities. Beyond a critical coupling and drive, the symmetry breaking states do not form. We analyze the occurrence and stability of the semiclassical solutions, finding that there can be up to nine solutions, a maximum of four of which are stable containing pairs of symmetry breaking and symmetry preserving states in a multistable region. We delineate a phase diagram for the symmetry broken states in the parameter space of drive and tunneling.

\begin{figure}[t]
\vspace{-0.0cm}
\begin{center}
\includegraphics[width=0.35\textwidth,angle=0]{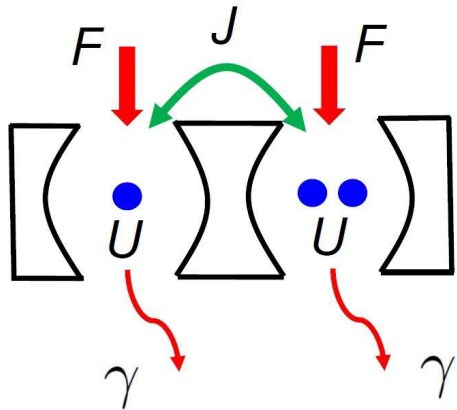}
\end{center}
\vspace{-0.6cm}
\caption{(Color online) Schematic for our model system of two coupled photonic cavities. Effective photon-photon interactions within each cavity give rise to a Kerr nonlinearity with strength $U$ and inter-cavity mode overlap contributes to tunneling with strength $J$. The system is coupled to a Markovian bath and photons can decay or leak out with rate $\gamma$. Coherent pumping ($F$) replenishes the photons. We treat the case when drive and dissipation are the same for both cavities. Two coupled cavities are the simplest setting for an array of coupled cavities where the interplay of dissipation, drive and interaction plays a crucial role that are being investigated in a wider context.}
\label{fig:schematic}
\end{figure}

We then study the system by solving the full quantum mechanical master equation and using the method of quantum trajectories~\cite{dalibard92,dum92,carmichael93,daley14}. In a full quantum treatment, when quantum fluctuations are taken into account, the symmetry breaking states are no longer seen, similar to the case of single-cavity bistable states~\cite{drummond80}. However, in quantum trajectory simulations of the dynamics, we show that quantum jump statistics of number differences reveal the presence of underlying semiclassical bistability indicating the presence of symmetry breaking states. Finally, we present a set of analytical expressions for the steady state correlation functions using the complex P-representation~\cite{drummond80,gardiner80} expressing the master equation in the form of Fokker-Planck equation. The solutions work for small tunnel couplings; we discuss the regime of validity of the solutions in the parameter space.

The article is organized as follows. In section I, we introduce the model. In section II, we present semiclassical analysis and analyze the driven-dissipative symmetry breaking states. In section III, we present quantum trajectory analysis and full quantum treatment of the master equation. In section IV, we derive analytical solutions for the steady state correlation functions, and summarize our results in section V.

\section{Model}

We consider two cavities with Kerr nonlinearity~\cite{drummond80} that are coupled by tunneling~\cite{hamel14,houck14}. The cavities are coherently driven in a dissipative setting, with both drive and dissipation acting equally on both sites. Fig.~\ref{fig:schematic} shows a schematic of the set up. The system can be described by the following Hamiltonian
\begin{eqnarray}
{\hat H} &=& -J ({\hat a}_1^{\dagger} {\hat a}_2+{\hat a}_2^{\dagger} {\hat a}_1) + \frac{U}{2} ({\hat a}_{1}^{\dagger2}{\hat a}_{1}^{2}+{\hat a}_{2}^{\dagger2}{\hat a}_{2}^{2})+\Delta \omega \nonumber\\
&& \times ({\hat a}_1^{\dagger} {\hat a}_1+{\hat a}_2^{\dagger} {\hat a}_2) + F ({\hat a}_1^{\dagger} +{\hat a}_2^{\dagger})+F^{*}({\hat a}_1+{\hat a}_2),
\end{eqnarray}
where ${\hat a}_1$ and ${\hat a}_2$ are the annihilation operators for the two cavities labelled $1$ and $2$. Here $J$ is the inter-cavity tunneling strength, $U$ is the anharmonicity or the nonlinear Kerr-type interaction strength, $F$ is the driving field for cavity 1 and 2, and $\Delta \omega=\omega_c-\omega_F$ is the detuning, after the rotating wave approximation for the drive term. $\omega_c$ and $\omega_F$ are the cavity resonance frequency and driving frequency, respectively. Tunnel couplings occur due to the overlap of the spatial profile of the cavity modes, and can be engineered in large arrays of resonators, e.g. in silicon rings ~\cite{vlasov07,hafezi13}, photonic crystal cavities~\cite{vuckovic12}, and exciton-polariton systems~\cite{bloch16}.


\begin{figure}[t]
\vspace{-0.0cm}
\begin{center}
\includegraphics[width=0.48\textwidth,angle=0]{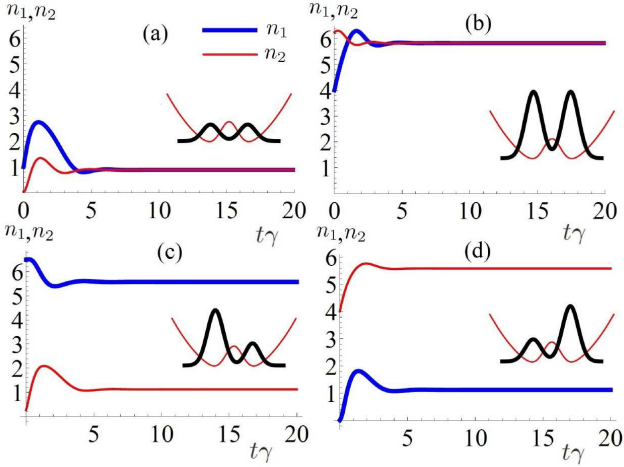}
\end{center}
\vspace{-0.4cm}
\caption{(Color online) Semiclassical dynamics to reach the steady states in a Bose-Hubbard dimer. For our model, there are non-equilibrium steady states where more photons are localized in one well than the other, a driven-dissipative analog of the symmetry breaking states of a closed system double well. Panels (a) and (b) show dynamics where the steady states have equal number of photons in the two cavities -- (a) low density state and (b) high density state. (c) and (d) show how states with unequal numbers of photons in the two cavities are reached. Symmetry breaking states come in pairs and are always in a multistable regime. Parameters used: $F/\gamma=2.6, J/\gamma=0.1, U/\gamma=0.6, \Delta \omega/\gamma=-3$. The four steady states are reached with four different initial values of $(n_1,\theta_1,n_2,\theta_2)$.}
\label{fig:dynamics}
\end{figure}

\begin{figure*}[ht]
\vspace{-0.5cm}
\begin{center}
\includegraphics[width=0.92\textwidth,angle=0]{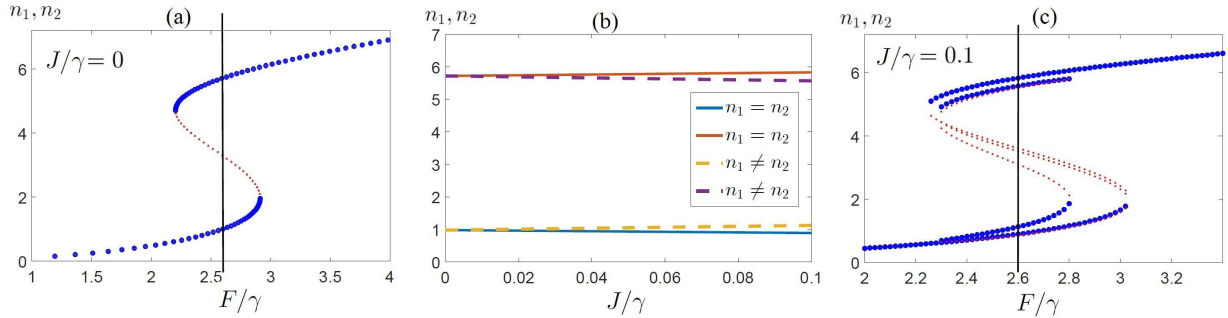}
\end{center}
\vspace{-0.5cm}
\caption{(Color online) From single cavity bistability to dimer multistability and driven-dissipative symmetry breaking states. (a) shows the semiclassical solutions in the limit of a single cavity when there is no coupling ($J=0$). Blue circles and red dots denotes stable and unstable states, respectively. For a fixed drive $F/\gamma=2.6$ (denoted by a vertical line in (a) and (c)), (b) depicts the change in the steady state photon occupations as the coupling $J/\gamma$ slowly increases -- four solutions originate from the two branches of the single-cavity solutions. (c) depicts all the solutions, both stable and unstable, which there are at most nine in some regime and a maximum of four of which are stable. For $F/\gamma=2.6$, we enter a regime of multiple solutions, two symmetry preserving states and a pair of symmetry breaking states. Same parameters are used as in Fig. 2.}
\label{fig:sbstates}
\end{figure*}

\begin{figure}[ht]
\vspace{-0.0cm}
\begin{center}
\includegraphics[width=0.37\textwidth,angle=0]{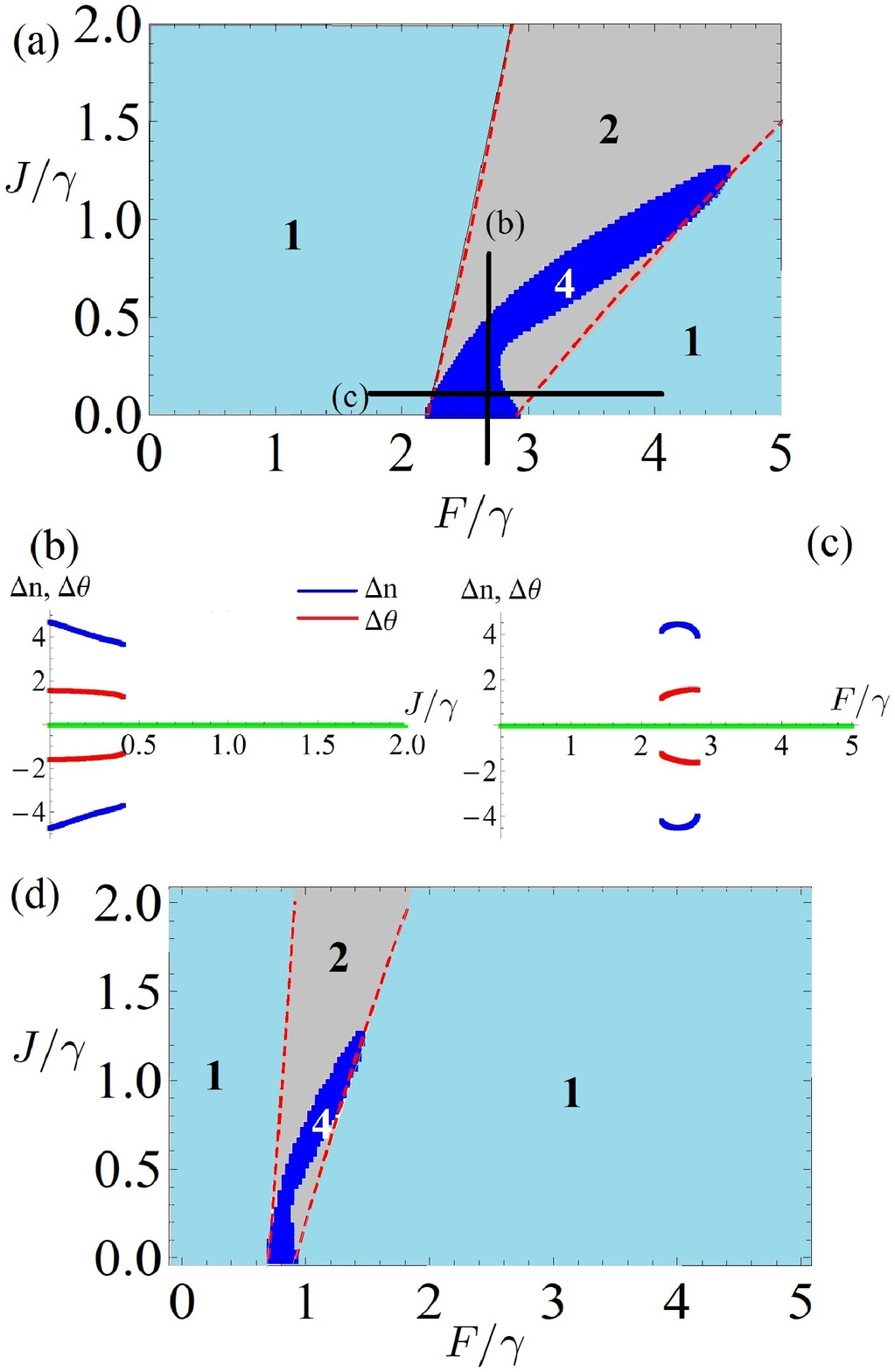}
\end{center}
\vspace{-1.3cm}
\begin{center}
\includegraphics[width=0.45\textwidth,angle=0]{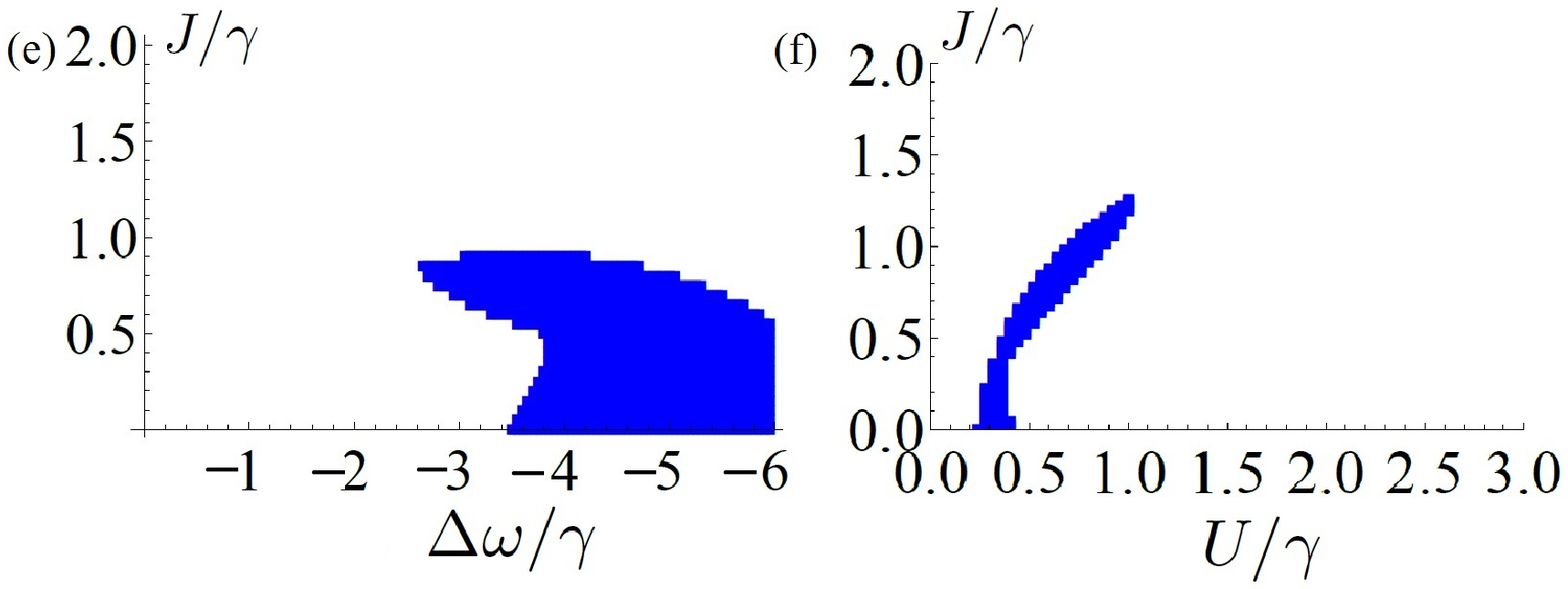}
\end{center}
\vspace{-0.7cm}
\caption{(Color online) Phase diagram for symmetry breaking states. (a) and (d) show phase diagram showing the presence of symmetry breaking states (the region labelled 4) in the tunneling-drive ($J/\gamma-F/\gamma$) plane for $\Delta \omega/\gamma=-3$ and different values of $U/\gamma=0.6$, and $6$ respectively. The area labelled 1 has one symmetry preserving steady state, 2 has two symmetry preserving states in a bistable region while 4 has two symmetry breaking and two symmetry preserving states. (b) shows $\Delta n$ and $\Delta \theta$ as a function of $J$ for a fixed $F=2.6$. (c) shows $\Delta n$ and $\Delta \theta$ as a function of $F/\gamma$ for a fixed $J/\gamma=0.1$. These depict how the symmetry breaking states appear and disappear as a function of tunnelling and drive. We see that there is a critical $J/\gamma$ value beyond which the symmetry breaking states do not appear. To gain a better insight, (e) and (f) depict the phase diagram in the tunneling-detuning ($J/\gamma-\Delta \Omega/\gamma$) and tunneling-interaction ($J/\gamma-U/\gamma$) planes, for fixed $F/\gamma=3.5, U/\gamma=0.6$ and $F/\gamma=3.5, \Delta \Omega/\gamma=-3$, respectively. The shaded regions contain symmetry breaking states.}
\label{fig:phasediag}
\end{figure}

For our open system, there can be photon losses induced by spontaneous decay or cavity leakage. In the approximation that the system is weakly coupled to a Markovian bath, the dynamics of the density matrix can be modeled by a quantum master equation in the Lindblad form
\begin{eqnarray}
\frac{\partial {\hat \rho}}{\partial t}&=&-i[{\hat H},\rho]+\gamma [2 ({\hat a}_1 {\hat \rho} {\hat a}_1^{\dagger}+{\hat a}_2 {\hat \rho} {\hat a}_2^{\dagger}) \nonumber \\
&&-({\hat a}_1^{\dagger} {\hat a}_1+{\hat a}_2^{\dagger}{\hat a}_2) {\hat \rho} - {\hat \rho}({\hat a}_1^{\dagger} {\hat a}_1+{\hat a}_2^{\dagger} {\hat a}_2)],
\end{eqnarray}
where $\gamma$ is the dissipation rate in each cavity. We set $\hbar=1$ throughout.

\section{Semiclassical analysis}

Ignoring quantum fluctuations, the mean field amplitudes for the field operators are
$\alpha_1=\langle \hat{a}_1 \rangle, \alpha_2=\langle \hat{a}_2 \rangle$. We assume $\alpha_1=\sqrt{n_1} e^{i \theta_1}$ and $\alpha_2=\sqrt{n_2} e^{i \theta_2}$, where $n_1, n_2$ are occupation numbers and $\theta_1, \theta_2$ are phases for the two cavities, and $\Delta n=n_1-n_2, \Delta \theta=\theta_1-\theta_2$. In the semiclassical approximation the correlation functions factorize, and the equations of motion become
\begin{eqnarray}
\frac{\partial}{\partial t} \alpha_1=F-\alpha_1 f(\alpha_1 \alpha^*_1)+i J \alpha_2 \nonumber\\
\frac{\partial}{\partial t} \alpha^*_1=F^*-\alpha^*_1 f^*(\alpha_1 \alpha^*_1)-i J \alpha^*_2 \nonumber\\
\frac{\partial}{\partial t} \alpha_2=F-\alpha_2 f(\alpha_2 \alpha^*_2)+i J \alpha_1 \nonumber\\
\frac{\partial}{\partial t} \alpha^*_2=F^*-\alpha^*_2 f^*(\alpha_2 \alpha^*_2)-i J \alpha^*_1.
\label{eq:eqofmotion}
\end{eqnarray}
Here $f(\alpha_1 \alpha^*_1)=\kappa+i U \alpha_1 \alpha^*_1$ and $\kappa=\gamma+i \Delta \omega$.

We find the steady states of these equations by solving the four coupled differential equations Eq.~(\ref{eq:eqofmotion}), examining the long-time dynamics for different sets of initial conditions. For most values of parameters we get a single steady state. However, for the value of parameters $F/\gamma=2.6, J/\gamma=0.1, U/\gamma=0.6, \Delta \omega/\gamma=-3$, we get four different steady states as depicted in Fig.~\ref{fig:dynamics} showing their long time dynamics. We take $\gamma=1$ for all our calculations in this article, essentially giving other parameters in units of dissipation. Two of the steady states have equal number of photon occupations in the two cavities as in Figs.~\ref{fig:dynamics}(a) and (b), one with low occupations and the other with higher occupations. We refer to these states as {\it symmetry preserving states}. In addition to these, we get steady states where the photon occupations are different in the two cavities as shown in Figs.~\ref{fig:dynamics}(c) and (d). This pair of states are asymmetric, mirror images of each other, and localized more in one of the cavities. We refer to these states as {\it symmetry breaking states}. These states are the driven-dissipative analog of the symmetry breaking states in Josephson coupled junctions which have been observed with ultracold atoms and photons in closed systems~\cite{oberthaler05,smerzi97,jbloch13}. Unlike the single cavity semiclassical solutions, the phases here are important as the relative phase between the two cavities cannot be gauged away. For the symmetry preserving states we find $\Delta \theta=0$, and for the symmetry breaking states, we have $\Delta \theta \neq 0$.

Alternatively, we set $\frac{\partial}{\partial t} \alpha=0$ to find the steady states:
\begin{eqnarray}
0=F-\alpha_1 f(\alpha_1 \alpha^*_1)+i J \alpha_2 \nonumber\\
0=F^*-\alpha^*_1 f^*(\alpha_1 \alpha^*_1)-i J \alpha^*_2 \nonumber\\
0=F-\alpha_2 f(\alpha_2 \alpha^*_2)+i J \alpha_1 \nonumber\\
0=F^*-\alpha^*_2 f^*(\alpha_2 \alpha^*_2)-i J \alpha^*_1.
\label{eq:eqsteadystate}
\end{eqnarray}
Here we have four equations with four unknowns $n_1,n_2,\theta_1$ and $\theta_2$ that are real. For certain parameter values, there can be more than one real solutions, up to a maximum of nine, out of which a maximum of four are stable. This is the regime of multistability. The stable solutions that we obtain with this method match with what we obtain from the previous method by looking at the long-term dynamics starting with different initial conditions.

We find stability by introducing small fluctuations to the steady state, obtaining linearized equations for the fluctuations and examining their eigenvalues. Introducing small fluctuations about the steady state
$\alpha_1(t)=\alpha_1+{\tilde \alpha}_1(t)$, $\alpha_2(t)=\alpha_2+{\tilde \alpha}_2(t)$,
\begin{eqnarray}
\frac{\partial}{\partial t}\begin{pmatrix}
{\tilde \alpha}_1(t)\\
{\tilde \alpha}_1^*(t)\\
{\tilde \alpha}_2(t)\\
{\tilde \alpha}_2^*(t)
\end{pmatrix}
=-\vec{A}\begin{pmatrix}
{\tilde \alpha}_1(t)\\
{\tilde \alpha}_1^*(t)\\
{\tilde \alpha}_2(t)\\
{\tilde \alpha}_2^*(t).
\end{pmatrix}
\end{eqnarray}
Here the stability matrix $\vec{A}$ is following:
\begin{eqnarray}
\vec{A}=
\begin{pmatrix}
\kappa +i 2 n_1 U & i \alpha_1^2 U & iJ & 0\\
-i \alpha_1^{*2} U & \kappa^* -i 2 n_1 U & 0 & -iJ\\
-iJ & 0 & \kappa +i 2 n_2 U & i \alpha_2^2 U\\
0 & iJ & -i \alpha_2^{*2} U & \kappa^* -i 2 n_2 U
\end{pmatrix}
\end{eqnarray}
where $\kappa=\gamma+i\Delta \omega$. When the real part of the eigenvalues of $\vec{A}$ are positive the solution is stable. Furthermore, we look at the determinant and trace, as done in Ref.~\cite{drummond80}, for which the Hurwitz criterion for stability requires that the trace and determinant of this matrix is nonzero and positive for stable eigenvalues.

\begin{figure*}[ht]
\vspace{-0.5cm}
\begin{center}
\includegraphics[width=0.85\textwidth,angle=0]{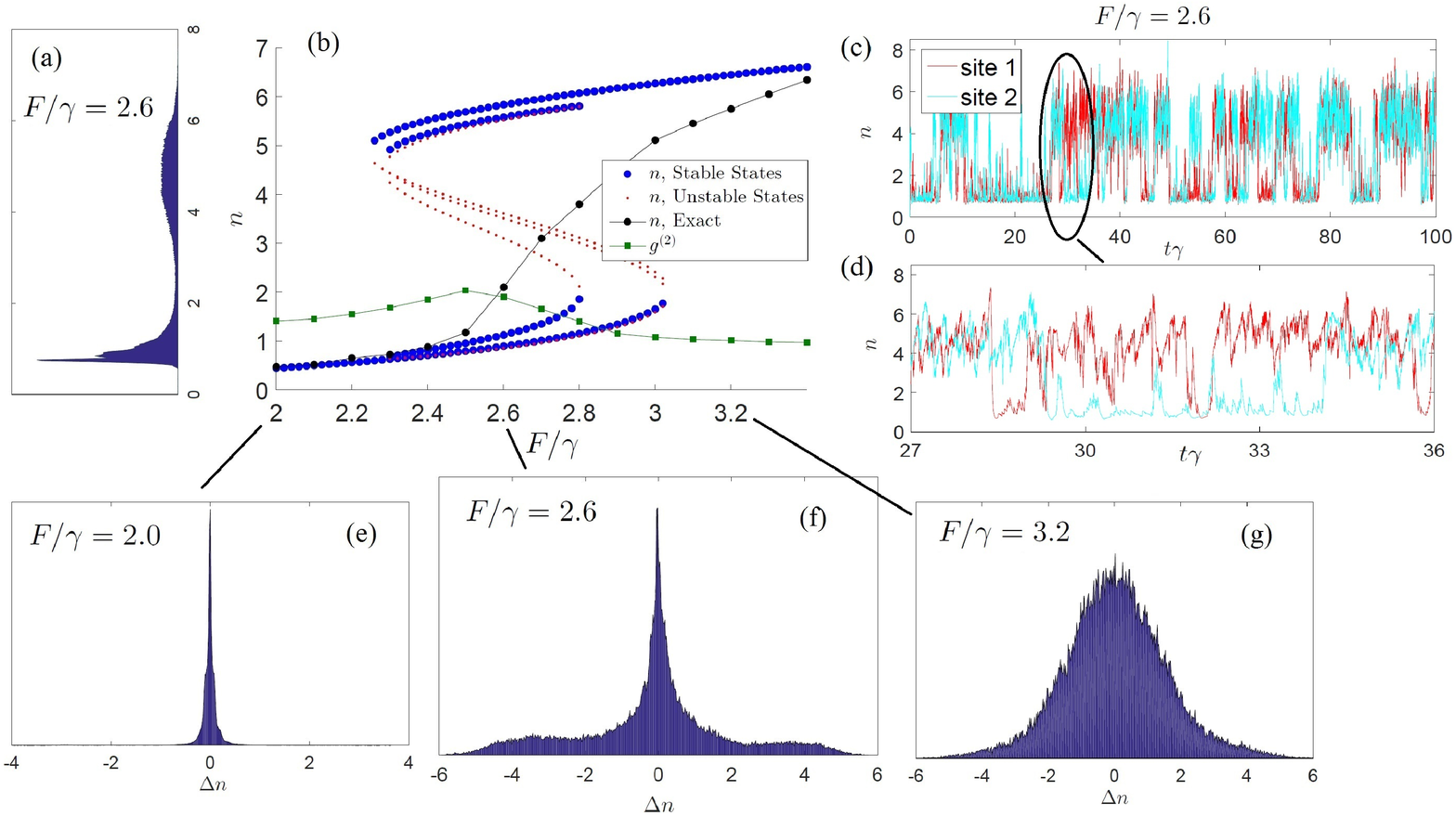}
\end{center}
\vspace{-0.5cm}
\caption{(Color online) Analysis of quantum trajectory simulations. (a) depicts a histogram of quantum jump statistics for photon occupations for cavity 1 for a sample trajectory in the bistable regime when $F/\gamma=2.6$, plotted along the $y$-axis of (b) which gives the multistability diagram for the same parameters as in Fig. 3(c). (b) also shows exact quantum steady state values of $n$ and $g^{(2)}$ obtained from the Lindblad master equation, showing unique values in the multistable regime and an enhancement of $g^{(2)}$ due to quantum fluctuations. (c) and (d) show the actual dynamics of $n_1$ and $n_2$ for $F/\gamma=2.6$ showing a sample quantum trajectory run. The magnified region in (d) shows that the two cavities can have unequal population during the dynamics which is enhanced when symmetry breaking states exist. In the space of photon number difference ($\Delta n$), histograms are depicted for (e) $F/\gamma=2$, (f) $F/\gamma=2.6$ and (g) $F/\gamma=3.2$ where the vertical axes are shown in arbitrary scale. (f) shows enhanced population difference described by a Lorentzian with broad side peaks indicating underlying symmetry breaking states. On the other hand, (e) and (g) is in regions outside multistability and have single central peaks. The number difference statistics thus contain signatures of underlying symmetry breaking bistability.}
\label{fig:trajectory}
\end{figure*}

We can further analyze the steady state equations Eq.~(\ref{eq:eqsteadystate}) to get two state equations
\begin{eqnarray}
|F|^2&=&n_1 (\gamma^2+(U n_1+\Delta \omega)^2) - \gamma J \sqrt{n_1 n_2} \sin \Delta \theta\nonumber\\
&& -2 J \sqrt{n_1 n_2} (U n_1+\Delta \omega)  \cos \Delta \theta +J^2 n_2 \nonumber\\
|F|^2&=&n_2 (\gamma^2+(U n_2+\Delta \omega)^2) + \gamma J \sqrt{n_1 n_2} \sin \Delta \theta \nonumber\\
&& -2 J \sqrt{n_1 n_2} (U n_2+\Delta \omega)  \cos \Delta \theta +J^2 n_1.
\label{eq:eqdrive}
\end{eqnarray}
Eqs.~(\ref{eq:eqdrive}) capture the parameter dependence for both the symmetry preserving and symmetry breaking semiclassical steady states. For symmetry preserving states, when $n_1=n_2=n$, $\Delta \theta=0$ and $J \neq 0$, this simplifies to one equation
\begin{equation}
|F|^2=n (\gamma^2 +(U n+\Delta \omega)^2) - 2 J n (U n+\Delta \omega)+J^2 n.
\label{eq:coupledbistability}
\end{equation}
Further taking the limit $J=0$ reproduces the single-cavity result of Drummond and Walls~\cite{drummond80}
\begin{equation}
|F|^2=n (\gamma^2 +(U n+\Delta \omega)^2).
\label{eq:singlecavity}
\end{equation}
Eq.~(\ref{eq:singlecavity}) is the bistability state equation for a single cavity. The extension of this to a coupled cavity gives us the state equations Eqs.~(\ref{eq:eqdrive}) and \ref{eq:coupledbistability}, expressing the dependence of $J$ in going from a single-cavity bistability to the coupled cavity bistability and multistability.

The symmetry breaking states are borne out of the bistable behavior of a single cavity. We can understand this by introducing an infinitesimal coupling to the two cavities as illustrated in Fig~\ref{fig:sbstates}. As soon as we turn on $J$, two bistable branches split into four steady states whose number occupations suggest that they are made up of the $low$ and $high$ density bistable states of individual cavities in this fashion - $(low,low)$, $(high,high)$, $(low,high)$, $(high,low)$. The $(low,high)$ and $(high,low)$ states are the symmetry breaking states. For a single cavity, there are three solutions to the semiclassical equations; two are stable and one is unstable. Coupling the two cavities gives rise to a maximum of nine solutions in some regime; out of which a maximum of four are stable. There are regions of one, two or four stable solutions. The regime of four stable solutions is the multistable region where two are symmetry preserving states and two are symmetry breaking states. Fig.~\ref{fig:sbstates}(c) shows both stable and unstable branch of the solutions marked in blue circles and red dots respectively. A pair of symmetry breaking states never appear in isolation but always come with a pair of symmetric states.

In Fig.~\ref{fig:phasediag} we analyze the steady state number and phase differences between the two cavities and delineate a phase diagram in the $J/\gamma-F/\gamma$ phase. Fig.~\ref{fig:phasediag}(b) shows the number difference $\Delta n$ as a function of $J/\gamma$ for a fixed $F/\gamma$, and we see that the self-trapped states appear as soon as we turn on $J$ and disappears for a critical $J/\gamma$. Here the positive and negative values of $\Delta n$ represent the pair of symmetry breaking states while the symmetric states are on zero of the $y$-axis. Similarly, Fig.~\ref{fig:phasediag}(c) shows the dependence on $F/\gamma$ for a fixed $J/\gamma$. With many of such slices for fixed $F/\gamma$ and $J/\gamma$, we can now draw a phase diagram for the presence of symmetry broken states in the tunneling-drive ($J/\gamma-F/\gamma$) plane for $\Delta \omega/\gamma=-3$ and different interactions (a) $U/\gamma=0.6$ and (d) $U/\gamma=6$. The shades differentiate three regions with one, two and four stable steady states. The region labelled $1$ has one stable symmetry preserving steady state. The region labelled $2$ is a bistable regime with two symmetry preserving states which continues to grow as $J/\gamma$ and $F/\gamma$ increases. The boundaries of the bistability region can be obtained from the analytic expression Eq.~(\ref{eq:coupledbistability}) which basically gives the turning points for the coupled cavity bistability in a similar way as can be derived for a single cavity~\cite{drummond80}. The boundaries of bistability region are shown in Figs.~\ref{fig:phasediag}(a), (d), having a near linear relationship in $F/\gamma$ and $J/\gamma$. The region labelled 4 and shaded dark blue is the region of multistability with two symmetry breaking states in addition to two symmetry preserving states. The symmetry breaking region shrinks and straightens upward as we increase the nonlinearity $U/\gamma$ as shown in Fig.~\ref{fig:phasediag}(d). The characteristics of this region are intimately connected to the whole parameter space of detuning, nonlinearity and that of bistability. We note that the symmetry breaking region always stays inside the bistability region, and therefore cannot go beyond the bistable phase boundaries to the right, where the slope a higher slope as $U/\gamma$ increases. As a function of $J/\gamma$, the phase diagram shows a reentrant behavior as we can see in Fig.~\ref{fig:phasediag}(a) near $F/\gamma=3$. To gain a better understanding, we depict the phase diagrams in two other parameters; in the tunneling-detuning ($J/\gamma-\Delta \Omega/\gamma$) and tunneling-interaction ($J/\gamma-U/\gamma$) planes, for fixed $F/\gamma=3.5, U/\gamma=0.6$ (e) and $F/\gamma=3.5, \Delta \Omega/\gamma=-3$ (f), respectively. The symmetry breaking regions which are shaded can be thought of as a cut in the multidimensional space of $(J/\gamma,\Delta \Omega/\gamma,F/\gamma,U/\gamma)$. In Fig.~\ref{fig:phasediag}(e) at detuning $\Delta \Omega/\gamma=-3$, we see that the symmetry breaking region matches with that of Fig.~\ref{fig:phasediag}(a) at $F/\gamma=3.5$; but the exact form of dependence in detuning is only apparent from the full diagram. One common feature we see is that there is always a maximum value of $J/\gamma$, and a range of $\Delta \Omega/\gamma, U/\gamma, F/\gamma$ values that confines the symmetry breaking regions.

In our model we do not break any symmetry of the system externally. Yet we get steady states which break the symmetry of the system and a stability diagram where there are multiple states for a single value of drive. For coherent pumping as we treat here, these driven dissipative symmetry breaking has not yet been observed. We note that the recent observation of spontaneous mirror symmetry breaking in coupled photonic-crystal nanolasers~\cite{hamel14} used incoherent pumping for which the mechanism and results are different in that the symmetry breaking states are not related to bistability as in our case. In terms of the bifurcation properties, unlike the supercritical pitchfork bifurfactions in experiment~\cite{hamel14}, we get here subcritical bifurcation~\cite{malomed15} as associated with bistable behavior. Pumping just one end cavity excites the system modes in a different way and gives rise to multistability~\cite{sarchi08}, which has been recently observed in photonic microcavities~\cite{bloch16}. Creating an external phase asymmetry such as in drive (driving two sites with $F$ and $-F$) or tunnelling (using $J$ instead of $-J$ in the Bose-Hubbard hamiltonian), also gives rise to symmetry broken states, whose criticality and entanglement has been investigated recently~\cite{ciuti16b}. Analysis from the perspectives of semiclassical discrete nonlinear Schr\"odinger equation (DNLS) for lattices and soliton physics have been reported in Ref.~\cite{naether15}.

\section{Quantum Analysis}

To understand more features of the driven dissipative Bose-Hubbard dimer, here we analyze the problem quantum mechanically taking into account the quantum fluctuations and using two methods -- first, we examine the dynamics by numerically solving the Lindblad master equation, and second, we do a quantum trajectory or Monte Carlo wavefunction analysis~\cite{dalibard92,dum92,carmichael93,daley14}.

In Fig.~\ref{fig:trajectory}(b) we show quantum steady state values overlaid on the semiclassical multistability diagram. For the quantum analysis, we use the Fock basis $|m_1,m_2\rangle$, where $m_1$ and $m_2$ are the occupations in cavities 1 and 2, respectively. A resulting equation of motion can now be constructed for the density matrix elements following the Lindblad master equation Eq.~(2). We now integrate the equation of motion using fourth-order Runge-Kutta to look at the long time evolution and determine the steady state. We find the quantum steady states to be unique for each value of the drive. Occupation ($n$) and normalized second-order correlator ($g^{(2)}$) are shown in Fig.~\ref{fig:trajectory}(b) as a function of drive, where $g^{(2)}=\frac{\langle a^{\dagger}a^{\dagger}aa \rangle}{n^2}$. In the multistability region, $g^{(2)}$ exhibits a peak which is due to the presence of enhanced quantum fluctuations that cause the underlying semiclassical multistability. This is similar to the case of single cavity bistability~\cite{drummond80}. However, the $g^{(2)}$ signatures only indicate that this is a region of bistability or multistability without giving us any clue about the presence of symmetry breaking states. To reveal this feature, we perform quantum trajectory analysis below.

Quantum trajectory method provides exact results for physical observables under the ensemble averaging of trajectories of wavefunctions. We specifically analyze quantum jumps in the multi-stable region and analyze signatures of the underlying semiclassical symmetry breaking solutions. First, we pick a specific driving field $F/\gamma=2.6$ to run the quantum trajectory simulations. Simulation results are analyzed in Fig.~\ref{fig:trajectory}: panel (a) shows a histogram of the quantum jumps as they happened as a function of the occupation numbers of one cavity plotted vertically along the semiclassical multistability diagram. The two peaks of the histogram coincide with the lower and upper branches of the multistability diagram as expected for bistability. Now if we investigate the dynamics of $n_1$ and $n_2$ separately, a typical such evolution for a single trajectory is given in Fig.~\ref{fig:trajectory}(c). This shows clearly that the photons spend most of the time fluctuating near the two bistable branches. Panel (d) shows a magnified region where the two cavities have unequal populations during the evolution. A histogram of jumps in the variable of photon population difference ($\Delta n=n_1-n_2$) for $F/\gamma=2.6$ is shown in panel (f). The distribution in (f) not only shows a single peak at $\Delta n=0$, but also broad side peaks at $\Delta n \neq 0$. For comparison, Fig.~\ref{fig:trajectory}(e) and (g) show the statistics for quantum jumps outside the multistable region at $F/\gamma=2.0$ and $F/\gamma=3.2$ respectively, showing single Lorentzian peaks at $\Delta n=0$. The highest peaks centered at $\Delta n=0$ correspond to the jumps related to symmetry preserving states, whereas, we interpret the broad side peaks of the distribution originating from the symmetry breaking states. In our model, the symmetry breaking states always coexist with the symmetric states. So the side peaks are always overshadowed by a prominent central peak. As for a quantum signature of driven-dissipative symmetry breaking states, we show here that a typical quantum trajectory reveals this feature. Recent experiments~\cite{mabuchi11} have used homodyne detection of photocurrents to detect the quantum trajectory underlying a bistability, and hence, the observation of quantum features of dimer symmetry breaking states should be within experimental reach.

Below we discuss some insights on the physical mechanism of multistability and symmetry breaking. Multistability in coupled cavities has the same origin as that of the bistability in a single cavity. In the semiclassical limit of a driven dissipative cavity, nonlinearity gives rise to two stable steady state solutions. In a quantum treatment, the quantum fluctuations leads to switching between these semiclassical states, and the density matrix is unique. We identify similar effects for two coupled cavities, as illustrated in this section. In coupled cavities, nonlinearity gives rise to multiple solutions corresponding to multiple minima in the steady state potential landscapes. However, the inclusion of quantum fluctuations leads to switching among multiple semiclassical solutions. More specifically, by analyzing the individual quantum trajectories, we find that photons jump between the stable branches in a way that reveals symmetry breaking in number difference.

In a closed system, symmetry breaking or self-trapped states appear due to nonlinearity getting larger than tunneling such that staying in one cavity minimizes the energy. In contrast, in an open driven-dissipative system, energy minimization does not determine the steady states, and we could not find a simple relation among the parameters of nonlinearity, tunneling, drive and detuning that could explain the symmetry breaking states. However, Eqs. (7) and (8), combined with the stability analysis, give relationships among the parameters for symmetry breaking and multistable states. We have presented phase diagrams for particular set of parameters, and one can understand the symmetry breaking states as a hybridization of unequal density bistable states from each individual cavity surviving in the coupled cavity limit for small coupling.

\section{Analytic solution for the steady state}

Analytic solutions for steady states of driven-dissipative systems are very rare. For a single cavity with Kerr nonlinearity, analytic solutions for the steady-state was presented by Drummond and Walls in Ref.~\cite{drummond80}. Since then, there have been a number of instances where analytic solutions have been obtained for various scenarios in a single-cavity, such as with two-body loss~\cite{ciuti16}. Drummond and Walls~\cite{drummond80} used the complex P-representation to obtain a Fokker-Planck equation and present closed-form analytic expressions for the correlation functions. We apply a similar technique here for the two-coupled nonlinear cavities and present analytic solutions for the correlation functions expressed as a series. The solutions are applicable for small $J$ and deviate slowly from the exact numerical results as $J$ is increased. Considering the fact that no analytic solutions are known for coupled cavities, we present our investigations here which may find applications in the study of weak coupling limit and in finding an improved solution. A short derivation is provided below.

In comparison to the single cavity case~\cite{drummond80}, the additional term in the dimer hamiltonian is the hopping term, $-J\,({\hat a}_{1}^{\text{\ensuremath{\dagger}}}{\hat a}_{2}+{\hat a}_{2}^{\dagger}{\hat a}_{1})$. After a rotating wave approximation, the contribution of this term to the master equation dynamics becomes: $iJ[{\hat a}_{1}^{\dagger}{\hat a}_{2}+{\hat a}_{2}^{\dagger}{\hat a}_{1},{\hat \rho}]$.

\begin{widetext}

Now the full master equation with Kerr nonlinearity is
\begin{eqnarray}
\dot{{\hat \rho}} & = & -i\Delta \omega [{\hat a}_{1}^{\dagger}{\hat a}_{1}+{\hat a}_{2}^{\dagger}{\hat a}_{2},{\hat \rho}]-i\frac{U}{2}[{\hat a}_{1}^{\dagger 2}{\hat a}_{1}^{2}+{\hat a}_{2}^{\dagger 2} {\hat a}_{2}^{2},{\hat \rho}]+[F {\hat a}_{1}^{\dagger}+F^{*} {\hat a}_{1},{\hat \rho}]+[F {\hat a}_{2}^{\dagger}+F^{*} {\hat a}_{2},{\hat \rho}]\nonumber\\
 &  & +iJ[{\hat a}_{1}^{\dagger}{\hat a}_{2}+{\hat a}_{2}^{\dagger}{\hat a}_{1},{\hat \rho}]+\gamma[2{\hat a}_{1}{\hat \rho} {\hat a}_{1}^{\dagger}-{\hat \rho} {\hat a}_{1}^{\dagger}{\hat a}_{1}-{\hat a}_{1}^{\dagger}{\hat a}_{1}{\hat \rho}]+\gamma [2{\hat a}_{2}{\hat \rho} {\hat a}_{2}^{\dagger}-{\hat \rho} {\hat a}_{2}^{\dagger}{\hat a}_{2}-{\hat a}_{2}^{\dagger}{\hat a}_{2}{\hat \rho}].
\end{eqnarray}
The contribution to the $P$ function dynamics from the hopping term is
\begin{eqnarray}
\int d\vec{\alpha}|{\alpha}\rangle \langle{\alpha}|\frac{dP(\vec{\alpha})}{dt}  =  \int d\vec{\alpha}(iJ({\hat a}_{1}^{\dagger}{\hat a}_{2}|{\alpha}\rangle \langle{\alpha}|+{\hat a}_{1}{\hat a}_{2}^{\dagger}|{\alpha}\rangle \langle{\alpha}|)-iJ(|{\alpha}\rangle \langle{\alpha}|{\hat a}_{1}^{\dagger}{\hat a}_{2}+|{\alpha}\rangle \langle{\alpha}|{\hat a}_{1}{\hat a}_{2}^{\dagger}))P(\vec{\alpha})\nonumber\\
  =  \int d\vec{\alpha}(iJ((\frac{\partial}{\partial\alpha_{1}}+\alpha_{1}^{\dagger})\alpha_{2}+(\frac{\partial}{\partial\alpha_{2}}+\alpha_{2}^{\dagger})\alpha_{1})-iJ((\frac{\partial}{\partial\alpha_{1}^{\dagger}}+\alpha_{1})\alpha_{2}^{\dagger}+(\frac{\partial}{\partial\alpha_{2}^{\dagger}}+\alpha_{2})\alpha_{1}^{\dagger}))|{\alpha}\rangle \langle{\alpha}|P(\vec{\alpha}).
\end{eqnarray}
where $\vec{\alpha}=(\alpha_1,\alpha_{1}^{\dagger},\alpha_2,\alpha_{2}^{\dagger})$.
Now the Fokker-Planck equation for the $P$ function is given as
\begin{equation}
\frac{\partial}{\partial t} P(\vec{\alpha})=(\partial_{\mu} A_{\mu}(\vec{\alpha})+\frac{1}{2} \partial_{\mu} \partial_{\nu} D_{\mu\nu}(\vec{\alpha})) P(\vec{\alpha}),
\end{equation}
where the drift matrix and the diffusion vector respectively are
\[
A_{\mu}=\begin{pmatrix}\kappa\alpha_{1}+iU\alpha_{1}^{2}\alpha_{1}^{\dagger}-F+iJ\alpha_{2}\\
\kappa^{*}\alpha_{1}^{\dagger}-iU^{*}\alpha_{1}^{\dagger 2}\alpha_{1}-F^{*}-iJ\alpha_{2}^{\dagger}\\
\kappa\alpha_{2}+iU\alpha_{2}^{2}\alpha_{2}^{\dagger}-F+iJ\alpha_{1}\\
\kappa^{*}\alpha_{2}^{\dagger}-iU\alpha_{2}^{\dagger 2}\alpha_{2}-F^{*}-iJ\alpha_{1}^{\dagger}
\end{pmatrix},\qquad D_{\mu\nu}=\begin{pmatrix}-iU\alpha_{1}^{2} & 0 & 0 & 0\\
0 & +iU\alpha_{1}^{\dagger 2} & 0 & 0\\
0 & 0 & -iU\alpha_{2}^{2} & 0\\
0 & 0 & 0 & +iU\alpha_{2}^{\dagger 2}.
\end{pmatrix},
\]
In order to have a specific form of analytic solutions for the steady states of the Fokker-Planck equation, the potential conditions must be satisfied~\cite{drummond80}. Now let us check whether the potential conditions, $\partial_\mu V_\nu=\partial_\nu V_\mu$, are satisfied for this multidimensional Fokker-Planck equation. Here
\begin{eqnarray}
V_{\rho} & = & (D_{\mu\nu})^{-1}(2A_{\nu}+\partial_{\sigma}D_{\nu\sigma})\\
 & = & \begin{pmatrix}(\frac{i2\kappa}{U}+2)\frac{1}{\alpha_{1}}-2\alpha_{1}^{\dagger}-\frac{i2F+2J\alpha_{2}}{U}\frac{1}{\alpha_{1}^{2}}\nonumber\\
-(\frac{i2\kappa^{*}}{U}-2)\frac{1}{\alpha_{1}^{\dagger}}-2\alpha_{1}+\frac{i2F^{*}-2J\alpha_{2}^{\dagger}}{U}\frac{1}{\alpha_{1}^{\dagger 2}}\\
+(\frac{i2\kappa}{U}+2)\frac{1}{\alpha_{2}}-2\alpha_{2}^{\dagger}-\frac{i2F+2J\alpha_{1}}{U}\frac{1}{\alpha_{2}^{2}}\\
-(\frac{i2\kappa^{*}}{U}-2)\frac{1}{\alpha_{2}^{\dagger}}-2\alpha_{2}+\frac{i2F^{*}-2J\alpha_{1}^{\dagger}}{U}\frac{1}{\alpha_{2}^{\dagger 2}}.
\end{pmatrix}
\end{eqnarray}
Denoting $\alpha_1, \alpha_1^{\dagger}, \alpha_2, \alpha_2^{\dagger}$ as $1,2,3,4$ respectively, we find $\partial_1 V_2=\partial_2 V_1=-2, \partial_3 V_4=\partial_4 V_3=-2, \partial_1 V_4=\partial_4 V_1=0$, and $\partial_2 V_3=\partial_3 V_2=0$. These four potential conditions are satisfied. However, two potential conditions are not satisfied: $\partial_1 V_3=-\frac{J}{U} \frac{1}{\alpha_2^2}, \partial_3 V_1=-\frac{J}{U} \frac{1}{\alpha_1^2}$ and $\partial_2 V_4=-\frac{J}{U} \frac{1}{\alpha_2^{\dagger 2}}, \partial_4 V_2=-\frac{J}{U} \frac{1}{\alpha_1^{\dagger 2}}$. Here $\partial_1 V_3=\partial_3 V_1$ if $\alpha_1=\alpha_2$ and $\partial_2 V_4=\partial_4 V_2$ if $\alpha_1^{\dagger}=\alpha_2^{\dagger}$. They are also approximately satisfied in the limit $J/U$ is small. We present the analytic solutions under this restricted condition.

The steady-state $P$ function is
\begin{eqnarray}
P_{ss} &=&  \exp(-\int V_{\rho}d\vec{\alpha})\nonumber \\
&=& \alpha_{1}^{(c-2)}\alpha_{1}^{\dagger(d-2)}\alpha_{2}^{(c-2)}\alpha_{2}^{\dagger(d-2)}\nonumber\\
&&\times \exp\left[-\frac{2}{U}\left(\frac{(iF+J\alpha_{2})}
{\alpha_{1}}+\frac{(iF+J\alpha_{1})}{\alpha_{2}}+\frac{(iF+J\alpha_{2})^{*}}{\alpha_{1}^{\dagger}}+\frac{(iF+J\alpha_{1})^{*}}
{\alpha_{2}^{\dagger}}\right)+2\alpha_{1}\alpha_{1}^{\dagger}
+2\alpha_{2}\alpha_{2}^{\dagger}\right].
\end{eqnarray}
We can rewrite the steady state $P$ function
\begin{eqnarray}
P_{ss} &=& \beta_{1}^{(2-c)}\beta_{1}^{\dagger(2-d)}\beta_{2}^{(2-c)}\beta_{2}^{\dagger(2-d)}\nonumber\\
&&\times \exp\left[-\frac{2}{U}\left((iF+\frac{J}{\beta_{2}})\beta_{1}+(i F+\frac{J}{\beta_{1}})\beta_{2}\right) -\frac{2}{U}\left((i F+\frac{J}{\beta_{2}})^{*}\beta_{1}^{\dagger}+(iF+\frac{J}{\beta_{1}})^{*}\beta_{2}^{\dagger}\right)
+\frac{2}{\beta_{1}\beta_{1}^{\dagger}}+\frac{2}{\beta_{2}\beta_{2}^{\dagger}}\right],
\end{eqnarray}
where $c=i 2\kappa/U,\;d=(i 2 \kappa/U)^{*},\;\kappa=\gamma+i\Delta \omega$, and $\beta_{1}=\frac{1}{\alpha_{1}},\beta_{1}^{\dagger}=\frac{1}{\alpha_{1}^{\dagger}},\beta_{2}=\frac{1}{\alpha_{2}},\beta_{2}^{\dagger}=\frac{1}{\alpha_{2}^{\dagger}}$.
We now get the analytic expression for any order of correlations. The zeroth order correlator or the normalization integral is
\begin{eqnarray}
I(c,d)  =  (2\pi)^{4}\sum_{n_{1},n_{2},n_{3},m_{1},m_{2},m_{3}}\frac{2^{n_{1}+m_{1}}}{n_{1}!n_{2}!n_{3}!m_{1}!m_{2}!m_{3}!}
\left(\frac{2J}{U}\right)^{n_2+n_3+m_2+m_3}\left(\frac{2F}{U}\right)^{2c+2d+2n_1+2m_1-4} \nonumber\\
\Gamma^{-1}(c+n_{1}+n_{3}-n_{2})\Gamma^{-1}(d+n_{1}+m_{3}-m_{2})\Gamma^{-1}(c+m_1+n_{2}-n_{3})\Gamma^{-1}(d+m_{1}+m_{2}-m_{3}).
\end{eqnarray}
First order on-site correlation function (occupation) is
\begin{eqnarray}
G^{(1)} = (2\pi)^{4}\sum_{n_{1},n_{2},n_{3},m_{1},m_{2},m_{3}}\frac{2^{n_{1}+m_{1}}}{n_{1}!n_{2}!n_{3}!m_{1}!m_{2}!m_{3}!}\left(\frac{2J}{U}\right)^{n_{2}+n_{3}+m_2+m_3} \left(\frac{2F}{U}\right)^{2c+2d+2n_1+2m_1-2}\nonumber\\
\Gamma^{-1}(c+n_{1}+n_{3}-n_{2}+1) \Gamma^{-1}(d+n_{1}+m_{3}-m_{2}+1) \Gamma^{-1}(c+m_1+n_{2}-n_{3}) \Gamma^{-1}(d+m_{1}+m_{2}-m_{3}).
\end{eqnarray}
And the second order on-site correlation function is
\begin{eqnarray}
G^{(2)} =  (2\pi)^{4}\sum_{n_{1},n_{2},n_{3},m_{1},m_{2},m_{3}}\frac{2^{n_{1}+m_{1}}}{n_{1}!n_{2}!n_{3}!m_{1}!m_{2}!m_{3}!}
\left(\frac{2J}{U}\right)^{n_{2}+n_{3}+m_2+m_3}\left(\frac{2F}{U}\right)^{2c+2d+2n_1+2m_1} \nonumber\\
\Gamma^{-1}(c+n_{1}+n_{3}-n_{2}+2) \Gamma^{-1}(d+n_{1}+m_{3}-m_{2}+2) \Gamma^{-1}(c+m_1+n_{2}-n_{3}) \Gamma^{-1}(d+m_{1}+m_{2}-m_{3}).
\end{eqnarray}

\end{widetext}

\begin{figure}[h]
\vspace{-0.0cm}
\begin{center}
\includegraphics[width=0.42\textwidth,angle=0]{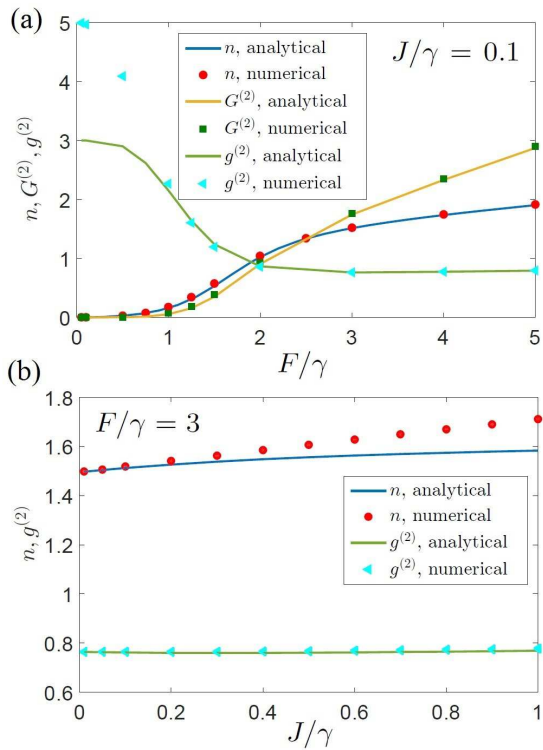}
\end{center}
\vspace{-0.6cm}
\caption{(Color online) A comparison of analytic and numeric solutions for parameters $U/\gamma=4$, $\Delta \omega/\gamma=-3$, $J/\gamma=0.1$. For these set of parameters, our analytic expressions give a very good match with the exact numerics for the first and second order correlation functions $n$ and $G^{(2)}$ shown in (a). For small drives $F/\gamma$, the normalized second order correlation $g^{(2)}$ however deviates. Please see the discussions in the texts for explanations. (b) shows comparisons for a fixed $F/\gamma=3$ and varying coupling $J/\gamma$. We find that for small values of $J/\gamma$ the solutions are comparable but deviates slowly as we go to higher $J/\gamma$.}
\label{fig:compare}
\end{figure}

A comparison of the results of the analytic solutions and numerical solutions of the master equation is shown in Fig.~\ref{fig:compare}. In Fig.~\ref{fig:compare}(a), for the set of parameters used, it is clear that the numeric and analytic results match very well for occupations $n$ and second-order correlation $G_{1}^{(2)}$. For small values of drive, however, the normalized second-order correlator $g^{(2)}=G_{1}^{(2)}/n^2$ shows a deviation. This is due to the fact that for this specific value of parameters, slight deviations in $G_{1}^{(2)}$ in the numerator and $n^2$ in the denominator enhances the discrepancies when $n$ is very small. The mismatch could also be due to the potential condition not satisfied in that region, as discussed earlier. In Fig.~\ref{fig:compare}(b) we show comparisons for increasing $J/\gamma$ for a fixed $F/\gamma$, and find deviations that come from the potential conditions not being satisfied. For increasing coupling the solutions deviate more and more, being comparable for small values of $J/\gamma$.

Unlike the single cavity case, the series here do not sum to a closed form hypergeometric function. Instead the computation involves a series involving six variables leading to difficulties in the convergence of the series. Due to the properties of the gamma function, sometimes the series converges fast. More specifically, $\Gamma(x)$ increases very fast with $x$ when $x$ is positive. Thus if we have the the detuning positive (i.e. the real part of $c$ and $d$ positive), the summation can converge for indices ($n_1,n_2,n_3,m_1,m_2,m_3$) up to around 10. If we have the detuning negative, we may need to compute up to a bigger index to make the series converge. We have checked the results for convergence for indices up to 30. For a smaller ratio of $J/U$ as in our example Fig.~\ref{fig:compare}, convergence is faster. It is conceivable that we can use Monte carlo sampling technique to perform this multidimensional sum.

A more fundamental issue is the regime of validity of the analytic expressions originating from the potential conditions as discussed earlier: $\partial_1 V_3=-\frac{J}{U} \frac{1}{\alpha_2^2}, \partial_3 V_1=-\frac{J}{U} \frac{1}{\alpha_1^2}$ and $\partial_2 V_4=-\frac{J}{U} \frac{1}{\alpha_2^{\dagger 2}}, \partial_4 V_2=-\frac{J}{U} \frac{1}{\alpha_1^{\dagger 2}}$. The potential conditions are approximately satisfied when $J/U$ is small. From our investigation we find that when $J/U$ is small, as shown in Fig.~\ref{fig:compare}, the analytical results are comparable to the numerical. The potential conditions are also satisfied when $\alpha_1=\alpha_2$, which involves both the occupation and phase in the two cavities to be equal. In the regime where quantum fluctuations are important such as in the bistability and the symmetry breaking region, $\alpha_1 \neq \alpha_2$ and the potential conditions are more likely to be violated. Analytical results are rare for a driven-dissipative system especially when going beyond a single cavity. The results presented here, although with a restricted regime of validity, are therefore important in our view. The analytic solutions presented here can be extended to multiple cavities.

\section{Conclusion}

We investigated the physics of two coupled nonlinear cavities in a lossy setting where both sites are driven coherently and equally. We performed semiclassical, quantum and analytical analyses of the system. In a semiclassical treatment, we find that the nonequilibrium steady states can have asymmetric number density in the two cavities which appear in addition to the symmetry preserving states. These states are the driven-dissipative analog of the double well self-trapped or symmetry broken states. Their appearance can be understood from the bistability of a single driven cavity; when two cavities are coupled, the low density and high density branches of single cavity bistable states hybridize to form two symmetry breaking states with unequal photon occupations in the two cavities. We examined the properties and stability of the semiclassical solutions, finding that there can be up to nine solutions of which a maximum of four are stable, giving rise to a pair of symmetry preserving and a pair of symmetry breaking states. We presented a phase diagram for these states in the tunneling-drive space.

We further studied the system using the method of quantum trajectories and by solving the full quantum mechanical master equation. In a full quantum treatment, when quantum fluctuations are taken into account, the coupled cavity bistable self-trapped states no longer appear, a case similar to that of single-cavity bistability. However, in a quantum trajectory analysis of the dynamics, we found that a histogram of quantum jumps in number differences reveal the presence of semiclassical bistability with strong indication of symmetry breaking states. Finally, we presented analytical solutions for the steady state correlation functions using the complex P-representation and forming a Fokker-Planck equation. We pointed out the regime of validity and limitations of this analytic solution.

Coupled cavity arrays are an exciting system to explore a host of important phenomena such as nonequilibrium dynamics, open system physics, strongly interacting photons and quantum many-body physics. We took the simplest coupled cavity model of a dimer and analyzed it using several different methods. The physics explored here and our predictions are within experimental reach in coupled cavity dimers, in terms of multistable states and single trajectory measurements. Besides fundamental physics, bistability and dimers have applications in optical memories and quantum correlation devices. The insights we gained on semiclassical and quantum nature of photons for two coupled cavities can also be useful for an array.


\begin{acknowledgments}
We thank Cristiano Ciuti, Howard Carmichael and Ryan Wilson for helpful discussions and exchanges.  We are grateful to Nicola Bartolo for sharing his derivations of analytic solutions. We acknowledge the support from AFOSR-MURI, ONR-YIP, ARO-MURI,  NSF-PFC at the JQI, and the Sloan Foundation. MH thanks the Kavli Institute of Theoretical Physics (KITP) at Santa Barbara for hospitality. KITP is supported by NSF PHY11-25915.
\end{acknowledgments}

\end{document}